\documentclass[10pt,prl,twocolumn,superscriptaddress]{revtex4}

\usepackage{graphicx}
\usepackage{amssymb}
\usepackage{times}
\usepackage{amsmath}
\usepackage{amsfonts}
\usepackage{txfonts}
\usepackage{color}
\usepackage{epstopdf}
\setcitestyle{super}

\begin{document}

\title{Ultrasensitive optical magnetometry at the microscale}

\author{S. Forstner}  \affiliation{School of Mathematics and Physics, University of Queensland, St Lucia, Queensland 4072, Australia}

\author{E. Sheridan} \affiliation{School of Mathematics and Physics, University of Queensland, St Lucia, Queensland 4072, Australia}

\author{J. Knittel} \affiliation{School of Mathematics and Physics, University of Queensland, St Lucia, Queensland 4072, Australia}\affiliation{Centre for Engineered Quantum Systems, University of Queensland, St Lucia, Brisbane, QLD 4072, Australia}

\author{C. L. Humphreys} \affiliation{School of Mathematics and Physics, University of Queensland, St Lucia, Queensland 4072, Australia}

\author{G. A. Brawley} \affiliation{School of Mathematics and Physics, University of Queensland, St Lucia, Queensland 4072, Australia}

\author{H. Rubinsztein-Dunlop} \affiliation{School of Mathematics and Physics, University of Queensland, St Lucia, Queensland 4072, Australia} \affiliation{Centre for Engineered Quantum Systems, University of Queensland, St Lucia, Brisbane, QLD 4072, Australia}

\author{W. P. Bowen}\affiliation{School of Mathematics and Physics, University of Queensland, St Lucia, Queensland 4072, Australia} \affiliation{Centre for Engineered Quantum Systems, University of Queensland, St Lucia, Brisbane, QLD 4072, Australia}

\date{\today} \maketitle
\textbf{Recent advances in optical magnetometry have achieved record sensitivity at both macro- and nano-scale. Combined with high bandwidth and non-cryogenic operation, this has enabled many applications. By comparison, microscale optical magnetometers have been constrained to sensitivities five orders-of-magnitude worse than the state-of-the-art. Here, we report an ambient optical micro-magnetometer operating for the first time in the picoTesla range, a more than three order-of-magnitude advance on previous results.
Unlike other ultrasensitive optical magnetometers, the device operates at earth field, achieves tens of MHz bandwidth, and is integrated and fiber coupled. Combined with 60~$\mu$m spatial resolution and microWatt optical power requirements, these unique capabilities open up a broad range of applications including cryogen-free and microfluidic magnetic resonance imaging, and electromagnetic interference-free investigation of spin physics in condensed matter systems such as semiconductors and ultracold atom clouds.}


Cryogenic SQUIDs are the favoured magnetometer for most high precision applications, combining excellent sensitivity with reliability. However, optical magnetometers are rapidly becoming an attractive alternative due to their non-cryogenic operation, ultrahigh sensitivity, low power requirements, and potential for high measurement bandwidth\cite{refBudker}. At milli-meter scale, atomic ensemble based magnetometers have achieved record sensitivities down to 160 aT~Hz$^{-1/2}$~~\cite{ref8}; while NV (nitrogen vacancy) diamond magnetometers provide the benchmark at nanoscale, with sensitivity in the range of several nT~Hz$^{-1/2}$~~\cite{ref16}. However, many applications in areas such as medical imaging\cite{refFagaly}, defense\cite{refRobbes}, solid state research\cite{refPannetier} and space science\cite{refAcuna}, benefit from spatial resolution in the intermediate microscale. At this scale, optical magnetometers have generally operated with sensitivity far worse than the state-of-the-art. Significant efforts have been made to miniaturize and microfabricate atomic vapour cell based magnetometers over the past decade. However, the smallest such devices remain constrained to 0.5 mm dimensions\cite{refBudker}. The only ultraprecise optical micro-magnetometers reported to date rely on ultracold atoms trapped in optical traps within bulky and restrictive ultrahigh vacuum systems\cite{refVengalattore, refmitchell}. 







\begin{figure}[t]
\centering
\includegraphics[width=9cm]{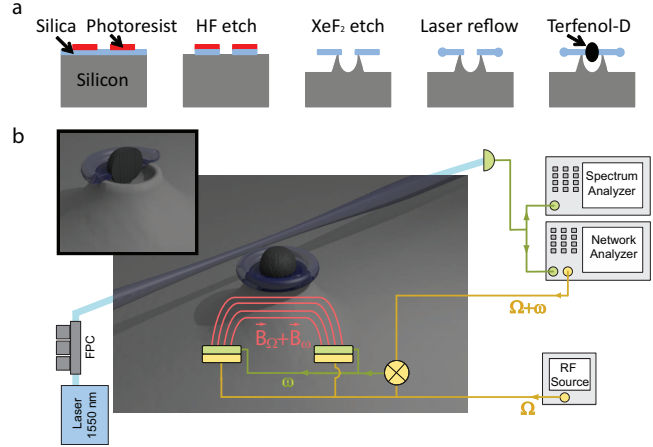}
\caption{{\bf Fabrication process and experimental schematic.} {\bf a}, Schematic representation (Side-view) of the magnetometer fabrication steps. {\bf b}, Schematic of experimental set-up. FPC: Fiber polarisation controller. Inset: 3D image showing internal structure of sensor.}
\label{fig: Layout}
\end{figure}

In this article we present a microscale cavity optomechanical magnetometer which uniquely combines picoTesla sensitivity, operation in ambient conditions, bandwidth in the tens of MHz, and the ability to operate at and above earth field. Furthermore, these results are achieved in an integrated all-optical ultralow power device.
A peak sensitivity of 200~pT~Hz$^{-1/2}$ is achieved, three orders of magnitude better than the previous best ambient optical magnetometer of similar size, with nonlinear frequency mixing providing nanoTesla sensitivity down to frequencies as low as 2~Hz, limited by the bandwidth of the measurement electronics rather than the sensor itself.  The six-order-of-magnitude dynamic range achieved is superior to the best NV\cite{ref16} and miniature SQUID magnetometers\cite{refBaudenbacher}; and significantly, exceeds the earth's field so that unshielded sensing can be performed. While the 40 MHz bandwidth surpasses the best previous optical micro-magnetometer by a factor of twenty five\cite{refShin}, and miniature-SQUIDs\cite{refBaudenbacher} by three orders of magnitude. 





Cavity optomechanical magnetometers were proposed recently\cite{refStefPRL}.
 The key principle is to combine microscale cavity optomechanics with magnetostrictive magnetometry. The combination of mechanical and optical resonances present in cavity optomechanical systems provide enhanced mechanical response to applied forces and optical readout with attometer precision\cite{refKipp,ref29}. In the presence of a magnetic field, the magnetostrictive material expands, exerting a measurable force upon the cavity optomechanical system. A proof-of-principle-experiment in Ref.~\cite{refStefPRL} demonstrated 400 nT~Hz$^{-1/2}$ sensitivity using a grain of magnetostrictive material bonded via epoxy to the surface of a conventional microtoroidal cavity optomechanical system. However, the bandwidth and sensitivity were both critically constrained by poor coupling of the magnetostrictive expansion to the mechanical resonances of the device. Furthermore, the sensor was unable to function in the Hz-kHz regime crucial to many applications, due to low frequency technical noise. Here, these constraints are overcome through the development of a new fabrication process allowing the magnetostrictive material to be embedded directly within the microtoroid structure, and by harnessing nonlinearities inherent in the magnetostrictive material to mix low frequency signals up to high frequency. 



Fig.~\ref{fig: Layout}a shows the fabrication process. A ring of photoresist is patterned on the top surface of a silicon wafer with a thermally grown silica top-layer (2 $\mu$m thick) using standard photolithographic methods. A buffered oxide (HF) etch is used to transfer the pattern into the silica layer. A xenon difluoride (XeF$_{2}$) etch undercuts the silica disk and etches a void in the supporting silicon pedestal. The silica disk is reflown using a CO$_{2}$ laser to produce a silica torus. Finally an appropriately sized piece of the magnetostrictive material Terfenol-D is positioned and epoxy bonded inside the central void. The final devices exhibit typical optical quality factors exceeding $10^6$, with multiple relatively broad mechanical modes ($Q\approx40$) providing broadband mechanical response over the frequency range from 5~to~40~MHz. 
This allows the fundamental thermal noise limit to micromechanical magnetometry to be reached, a limit typically precluded by several orders of magnitude in magnetostrictive magnetometers that rely on electrical read-out\cite{refKistenmacher}.

The sensitivity of the final magnetometer is characterized using the experimental setup shown in Fig.\ref{fig: Layout}b. For linear measurements of the high frequency sensitivity, the Terfenol-D is magnetized using a permanent bias magnet to maximize the linear magnetostrictive response\cite{refmagbook}. A pair of solenoids is then used to create a spatially uniform signal magnetic field across the device at radio frequencies. Light from a 1550 nm shot noise limited tunable fiber laser is guided through a polarization controller and evanescently coupled into the microtoroidal optical cavity via a tapered optical fiber. The laser frequency is thermally locked to the half maximum of an optical resonance. Strain in the resonator induced by the magnetostrictive material shifts the optical resonance frequency, thus modulating the amplitude of the transmitted light. This transmitted field is detected on an InGaAs photodiode, with only 50~$\mu$W of off resonant light required at the detector to achieve good signal-to-noise. Similar to Ref.~\cite{refStefPRL}, network/spectral analysis allow the magnetic field sensitivity to be determined as a function of signal frequency.


\begin{figure}[t]
\centering
\includegraphics[width=8cm]{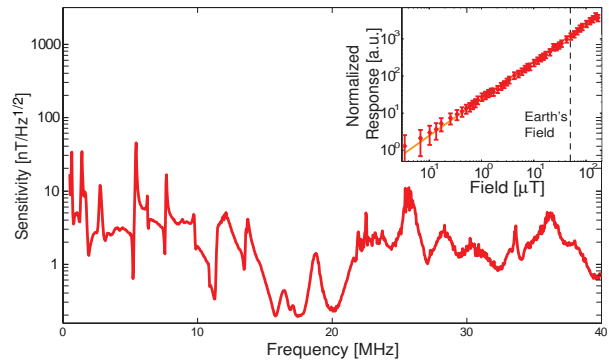}
\caption{{\bf High frequency sensitivity.} Sensitivity as a function of frequency in linear mode of operation, resolution bandwidth was 10 kHz. Inset: Sensor's response as a function of signal field strength at a resolution bandwidth of 100 Hz. The data was normalized by dividing by the noise floor. Dashed line: Earth's average magnetic field.}
\label{fig: HF}
\end{figure}
The measured magnetic field sensitivity of the device is shown in Fig.~\ref{fig: HF} over a range from 1 to 40 MHz. As can be seen, the sensitivity exceeds 10 nT~Hz$^{-1/2}$ over the majority of the range, with a peak sensitivity of 200 pT~Hz$^{-1/2}$ achieved at frequencies near 20~MHz, where the dominant mechanical modes are predominantly radial in nature and thus strongly coupled to the optical resonance frequency. This peak sensitivity outperforms previously reported ambient optical magnetometers\cite{refStefPRL} and electrically read-out magnetostrictive magnetometers\cite{refKistenmacher} of comparable size by more than three orders-of-magnitude. 
The detection bandwidth of 1 to 40 MHz shown in Fig.~\ref{fig: HF} is a 25-fold improvement on that achieved with the best previous NV magnetometer \cite{refShin}, and four orders of magnitude greater than the most sensitive atomic magnetometers \cite{refshah, ref8}. This high detection bandwidth is highly advantageous for applications such as microfluidic magnetic resonance imaging (MRI), where it enables observation of higher frequency spin-precession, and therefore improved sensitivity and resolution\cite{refufl_mri}. 

\begin{figure*}[t!]
\begin{center}
\includegraphics[width=18cm]{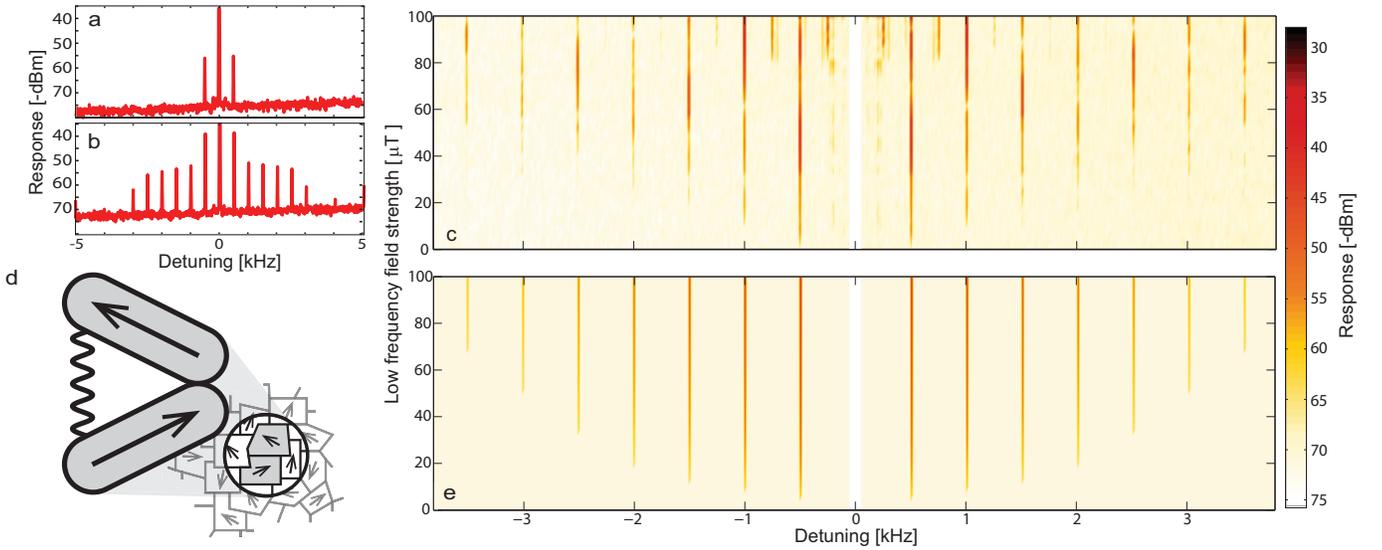} 
\caption{{\bf Response in nonlinear mode of operation.} {\bf a}, and {\bf b}, Spectra of sensor's response centred on the mixup frequency of 5.22 MHz under application of low frequency signals at 500 Hz with 1~$\mu$T and 22~$\mu$T amplitude, respectively. The resolution bandwidth was 10 Hz.
 {\bf c}, and {\bf e}, Pseudocolor plots of the measured and modelled sensor response spectrum as a function of signal strength and detuning from 5.22 MHz. Note: in these plots the strong linear response to the RF drive at frequencies close to zero detuning has been excluded. {\bf d}, Basic structure of magnetic domain pairs used to obtain simple microscopic model of mix-up process (see Supplementary Information for details). Model parameters: $\gamma/2\pi =160$~kHz, $\omega_m/2\pi = 5.2$~MHz , $\chi = 1.5 \times 10^{21}$~s$^{-2}$~m$^{-1}$, $c_0 = 3.1 \times 10^7$~m~s$^{-2}$~T$^{-1}$, $c_1 = 3.3 \times 10^{18}$~s$^{-2}$~T$^{-1}$, $c_2 = 8.2 \times 10^{25}$~s$^{-2}$~T$^{-1}$~m$^{-1}$, and $B_{\rm RF} = 17~\mu$T.}
\label{fig: pcolor}
\end{center}
\end{figure*}

The unique combination of high sensitivity, high bandwidth, and electromagnetic interference-free operation in a microscale device, could facilitate practical applications in areas such as imaging\cite{ref18}, material and circuit characterization\cite{refmagmat}, and medical diagnosis\cite{refWalker}. It may, further, enable a range of applications in fundamental science. For instance, there is much interest in condensed matter physics in the ability to detect single nuclear spins, and the interaction of pairs or ensembles of spins in semiconductor systems\cite{Zhao2012,GaAsspin}. Our devices may allow investigations of such physics with increased bandwidth and at a new range of frequencies. Similarly, strong spin-spin interactions in clouds of ultracold atoms allow the formation of spinor condensates, providing a unique new tool to study the behaviour of quantum gases\cite{refStenger}. However, direct laser based interrogation of the spin state is fundamentally limited by spontaneous emission\cite{refHope}. Since spinor condensates can exhibit MHz magnetic fields with strength in the range of hundreds picoTesla, our magnetometers provide a route to overcome this limit. Furthermore, since the size of our devices is comparable to a neuron, they may provide a new interference-free approach to mapping neuron magnetic fields\cite{refneuron,refneur2}, or indeed the magnetic fields of other comparably sized biological structures.




The dynamic range of the sensor was tested by varying the amplitude of the signal field over several orders of magnitude. As shown in Fig.~\ref{fig: HF} (inset), the response is linear over the full measurement range from 200 pT~Hz$^{-1/2}$ to 150 $\mu$T~Hz$^{-1/2}$, well above earth field and limited by the range of the testing coils rather than the sensor itself. To our knowledge, no previous ultrasensitive magnetometer of similar size has been capable of operating above earth field\cite{ref16, refShin, ref8}. 

Many significant applications require sensitivity to magnetic fields in the Hz to kHz frequency regime, including magnetic anomaly detection (MAD)\cite{ref6b}, geophysical surveys\cite{refgeo} and magnetoencephalography (MEG)\cite{refRomalis}. This regime is precluded in the linear mode of operation of our device due to low frequency technical noise from vibrations, temperature fluctuations, and electronic noise. Furthermore, the mechanical resonance frequencies of micrometer-scale structures, such as microtoroids, naturally lie in the MHz-range such that the mechanical response to the magnetostrictive force is intrinsically weaker at low frequencies. However, the magnetic domain structure of magnetostrictive materials make them intrinsically highly non-linear\cite{refmagbook}. This provides a mechanism through which to mix low frequency magnetic signals up to higher frequencies, and also has the effect of suppressing ambient low-frequency noise sources\cite{refDeganais}. Essentially, changes in the magnetic domain structure of the material due to low frequency fields alter the response of the material to an applied high frequency field\cite{refDeganais}. To explore such phenomena in cavity optomechanical magnetometers, a strong RF and weak Hz-kHz magnetic field are simultaneously applied to the device and the response analysed.  An RF source is used to produce an RF signal at frequency $\Omega$, while the network analyser produces a second RF signal at frequency $\Omega + \omega$, where $\omega$ is swept in frequency over a Hz-kHz range. The signal at $\Omega$ is split on an RF power splitter and applied to both a pair of air solenoids to generate the strong RF magnetic field, and to a mixer along with the signal at frequency $\Omega+\omega$, producing a low frequency signal at frequency $\omega$. The low frequency signal is applied to a second set of independent air solenoids to produce the low frequency magnetic field. Any nonlinear mixing between the low and high frequency fields caused by the nonlinear magnetostrictive response can be analysed, similarly to the case of linear operation, using spectral and network analysis. Note that electronic filtering was used to ensure that the detected signal did not arise from the electronic mix-up process, this was confirmed with control experiments where the magnetic field was verified using a pickup coil and a magnetoresistive sensor.

\begin{figure*}
\begin{center}
\includegraphics[width=18cm]{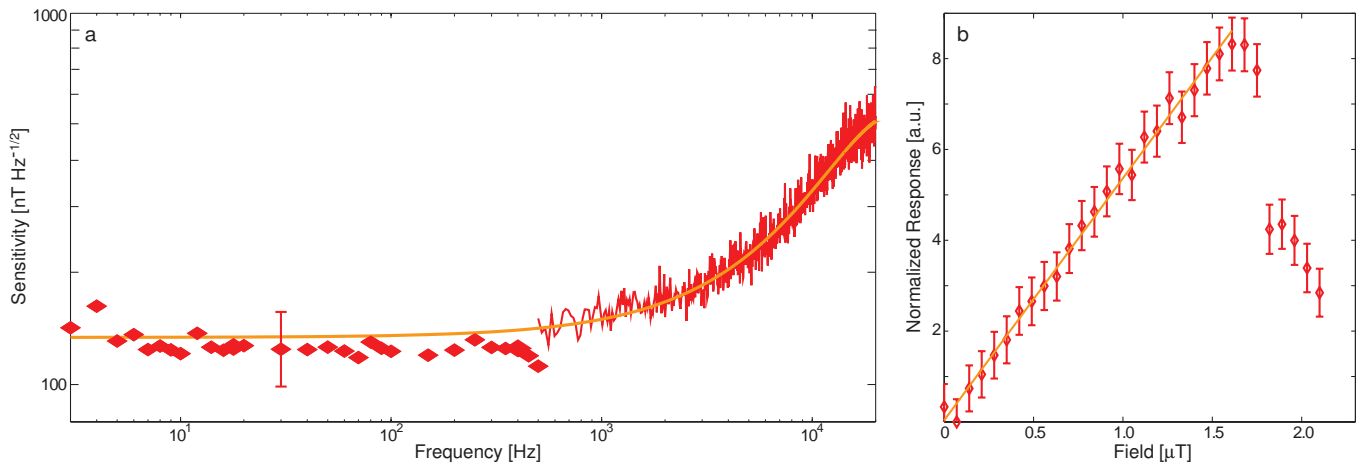}
\caption{{\bf Low frequency sensitivity.} {\bf a}, Low frequency sensitivity achieved by nonlinear mixup as a function of signal frequency. At frequencies above 500~Hz the sensitivity was determined using a combination of network and spectrum analysis at 10 Hz resolution bandwidth. At lower frequencies, the network response was determined individually from each power spectrum for each data point, at a resolution bandwidth of 1 Hz.
 The error bar shown for a single low frequency data-point is representative of the uncertainty of all points.
 {\bf b}, Mixup response as a function of low frequency signal field strength at a resolution bandwidth of 10 Hz. The data was normalized by dividing by the noise floor.}
\label{fig: LFMixup}
\end{center}
\end{figure*}

Fig.~\ref{fig: pcolor}a shows a typical optically detected power spectrum upon application of a strong RF drive field at $5.221$ MHz and a 1 $\mu$T signal field at 500 Hz. A pair of sidebands are evident, spaced 500 Hz either side of the $5.221$ MHz drive. This demonstrates that low frequency fields can be mixed-up to high frequencies in our architecture. Interestingly, as the strength of the low frequency field is increased above about 2 $\mu$T, a frequency comb of higher order sidebands appears, as shown in Fig. \ref{fig: pcolor}b-c. To our knowledge, such a frequency comb has not previously been reported in magnetostrictive magnetometers. To understand the appearance of this comb we developed a simple microscopic model of the interaction of a pair of domains within the Terfenol-D with the applied magnetic field, as illustrated in Fig.~\ref{fig: pcolor}d and discussed in detail in the supplementary information\cite{ref35}. This microscopic model results in an equation of motion for the dynamics of the domains, neglecting domain-domain interactions, given by
 \begin{equation}
\ddot x + \gamma \dot x + \omega_m^2 x + \chi x^2 = (c_0 - c_1 x - c_2 x^2) B \label{eqn}
\end{equation}
where $x$ is the offset of the length of the domain pair away from equilibrium, $\gamma$ is the dissipation rate of mechanical energy, $\omega_m$ is the mechanical resonance frequency due to the elasticity of the material, and $B = B_{\rm sig} \cos \omega t + B_{\rm RF} \cos \Omega t$ is the applied magnetic field with $B_{\rm sig}$ and $B_{\rm RF}$ being the amplitude of the low frequency signal field and radio frequency drive fields, respectively. The right-hand-side describes the forcing due to the magnetic field, and exhibits a linear term with coefficient $c_0$, responsible for our linear magnetometry results, and higher order nonlinear terms with coefficients $c_1$ and $c_2$. These higher order terms cause nonlinear mixing between magnetic fields at different frequencies, and have been used previously in magnetostrictive magnetometers to achieve low frequency sensitivity\cite{refDeganais}. Notably, however, our microscopic model predicts a second new form of nonlinearity, depending purely on the mechanical motion of the domains, with nonlinear co-efficient $\chi$ (see Eq.~(\ref{eqn})). It is this second nonlinearity that results in the frequency comb evident in our data. Using Eq.~(\ref{eqn}) to model the dynamics of our device yields the frequency response given in Fig.~\ref{fig: pcolor}e, which shows excellent qualitative agreement with experiment (Fig.~\ref{fig: pcolor}c). 

The sensitivity of the nonlinear mode of operation was determined in a similar way as that for the  linear mode of operation, but this time using the permanent bias magnet to maximize the nonlinear response of the device. Fig.~\ref{fig: LFMixup}a shows the achieved sensitivity at frequencies between 2 Hz and 20 kHz, with a peak sensitivity of 150 nT~Hz$^{-1/2}$ over the range 2 Hz to 1 kHz. The sensitivity deteriorates at higher frequencies, probably constrained by the bandwidth of the mechanical nonlinearity. Although not shown in the figure, the sensitivity also deteriorates at frequencies below 0.5 Hz, where we observe discrete steps in the magnetometer response attributed to Barkhausen noise from random domain flips in the Terfenol-D. Fig.\ref{fig: LFMixup}b shows the nonlinear response as a function of the strength of the applied signal field, for a signal frequency of 1 kHz. 
As can be seen, it is linear up to a threshold field strength of 1.7 $\mu$T, above which it drops precipitously. This drop corresponds to the onset of the frequency comb, with energy being transferred into higher order sidebands, and constrains the dynamic range in nonlinear operation. 

 

To conclude, we have developed an optical micro-magnetometer that achieves sensitivity in the picoTesla range. The magnetometer is all-optical, requires only microWatts of optical power, and has higher dynamic range and bandwidth than any previously reported precision microscale magnetometer. By leveraging inherent mechanical nonlinearities, we achieve nanoTesla range sensitivity at frequencies as low as 2 Hz. 
These capabilities provide the opportunity for new applications such as medical diagnosis with microscale samples using microfluidic nuclear magnetic resonance and microscale MRI\cite{refufl_mri}, and direct measurement of dynamics in strongly interacting spin systems such as semiconductors\cite{refWrachtrup}, superconductors\cite{refJang,refBouchard}, and spinor condensates\cite{refChapman, refStenger}. 
With further improvements in sensitivity possible\cite{refStefPRL}, these devices could extend the use of microscale magnetometers to areas including magnetic mapping of a single neuron grown on chip\cite{refHalgren} and low field MRI, allowing portable high resolution imaging at low power and low cost, and without the need for cryogenics.

\emph{Acknowledgements} This research was funded by the Australian Research Council Centre of Excellence CE110001013 and Discovery Project DP0987146 and by DARPA via a grant through the ARO. Device fabrication was undertaken within the Queensland Node of the Australian Nanofabrication Facility.

\end{document}